\documentclass[prl,twocolumn,superscriptaddress,letterpaper,lengthcheck]{revtex4-1}
\usepackage{amsfonts}
\usepackage{amsmath}
\usepackage{amssymb}
\usepackage{graphicx}
\usepackage{graphicx,amssymb,amsbsy,color}
\usepackage{bm}
\usepackage{bbold}

\begin{document}

\title{Quantum crystallography of Rydberg-dressed Bose gases on a square lattice}
\author{Che-hsiu Hsueh}
\affiliation{Department of Physics, Osaka City University, Sugimoto 3-3-138, Sumiyoshi-ku, Osaka 558-8585, Japan}
\affiliation{Department of Physics, National Taiwan Normal University, Taipei 11677, Taiwan}
\author{Wen-Chin Wu}
\email{wu@phy.ntnu.edu.tw}
\affiliation{Department of Physics, National Taiwan Normal University, Taipei 11677, Taiwan}
\author{Makoto Tsubota}
\email{tsubota@sci.osaka-cu.ac.jp}
\affiliation{Department of Physics, Osaka City University, Sugimoto 3-3-138, Sumiyoshi-ku, Osaka 558-8585, Japan}
\affiliation{The OCU Advanced Research Institute for Natural Science and Technology (OCARINA), Osaka, Japan}

\date{\today}

\begin{abstract}
We numerically investigate the quantum crystallographic phases of a Rydberg-dressed Bose gas loaded on a square lattice by using the mean-field Gross--Pitaevskii model. For a relatively weak lattice confinement, the phases of ground state undergo amorphism, polycrystal, and polymorphism following the increase of the blockade radius, and if the confinement is stronger, a single crystal with a specific filling factor will be formed. In order to distinctively characterize these phases, the structure function is also studied. In such an anisotropic system, we report that the first diagonal element of the superfluid-fraction tensor should be a measurable quantity, and an anisotropy parameter can be defined. In addition, for such crystallographic phases, the interaction potential can manifest where the grain boundaries appear.
\end{abstract}
\maketitle

The existence of a matter state that simultaneously possesses solid and superfluid natures, so-called supersolid, attracts both experimentalists and theorists.
The intuitional candidate systems for finding a supersolid are solid helium and Bose--Einstein condensates (BEC). In the former, one anticipates finding superfluidity in a solid and, in contrast, finding solidity in the latter system, which is regarded as a superfluid. An interaction with a soft core is regarded as the crucial factor to the formation of a supersolid\cite{PhysRevLett.72.2426,PhysRevLett.98.195301,PhysRevLett.99.135301,PhysRevB.77.054513,Sepúlveda2010,PhysRevLett.104.195302,PhysRevLett.105.135301,PhysRevA.83.021602,PhysRevB.84.094521,Boninsegni2012,1742-6596-400-1-012037,PhysRevA.86.013619,PhysRevB.86.060510,PhysRevLett.108.175301,PhysRevLett.108.265301,RevModPhys.84.759,PhysRevA.87.063604,PhysRevA.87.061602,PhysRevA.88.033618,PhysRevA.88.043646,Macrì2014,1367-2630-16-3-033038}, or otherwise a three-body interaction in a dipolar BEC\cite{nature16485,PhysRevLett.115.075303,PhysRevA.92.061603,PhysRevA.93.011604,PhysRevA.93.033644,PhysRevLett.116.215301}. Such a soft-core interaction can be engineered in clouds of cold atoms weakly coupling the Rydberg state to the ground state\cite{PhysRevLett.104.195302,0953-4075-43-15-155003,PhysRevLett.105.160404,PhysRevLett.105.135301}. As a supersoild may be observable in experiments with Rydberg-dressed alkali atoms, other quantum crystallographic states, such as superglass, are expected to be established. The superglass corresponding to a matter state that simultaneously possesses superfluidity and a frozen amorphous structure\cite{PhysRevLett.96.105301,PhysRevB.78.224306,PhysRevLett.103.215302,PhysRevLett.104.215301,PhysRevLett.109.157202,PhysRevB.83.094530,PhysRevB.85.104205,PhysRevLett.109.157202,PhysRevLett.116.135303,Fantoni2016}.

A decisive evidence to confirm the superfluid nature of a given quantum system is the measurement or calculation of the superfluid fraction $f_{s}$. In a perfect superfluid system, $f_{s}\rightarrow1$, whereas $f_{s}$ reduces from 1 when spatial modulation or dynamical fluctuation occurs, which suppresses the long-range phase coherence of superfluids. In a lattice system, it has been shown that the superfluid fraction is equal to the ratio of bare to effective band mass of the system, $f_{s} = m/m^{*}$\cite{book}. This indicates that the reduction of $f_{s}$ is compensated by the increase of the effective mass $m^{*}$. In a higher dimensional system, the effective mass or the superfluid fraction should be a tensor, leading to the following question: what is the measurable property that can emerge from the anisotropic superfluidity? Here, we propose that the diagonal element of the first effective mass tensor or the reciprocal effective mass tensor should be a probably measurable quantity by studying the response of the particle to an abruptly applied force\cite{PhysRevLett.112.170404}.

One method to form crystalline structures is to consider the anisotropy of interaction\cite{PhysRevLett.104.215301,PhysRevLett.109.157202,PhysRevB.85.104205}; another method is to consider the effect from external potentials, e.g., the disorder potential. For applications in strongly disordered environments, an insulating phase of interacting bosons known as Bose glass is obtained\cite{PhysRevB.40.546,PhysRevB.88.134206,PhysRevLett.110.100601,PhysRevLett.110.075304,PhysRevLett.111.160403,PhysRevA.90.031603,PhysRevB.90.125144,PhysRevLett.112.225301,PhysRevLett.113.095301,PhysRevA.91.031604,PhysRevA.91.031601,PhysRevB.92.180201,PhysRevLett.114.255701,PhysRevLett.114.155301,PhysRevLett.114.105303,1367-2630-18-1-015015}. In applications in lattice potentials, owing to the competition between the length scales of supersolid itself and the external potential or the competition between the interaction and the potential energies, there is a transition from an incommensurate to a commensurate density in one-dimensional cases\cite{PhysRevA.92.013634,PhysRevA.93.063605}. Although any continuous-space supersolid is compressible\cite{PhysRevLett.94.155302}, the extent of the compression is small. Owing to the lack of adjustment of the distance between supersolid droplets, in 1D cases, modulating the density may be the only possible way to reduce the raised potential energy. Comparatively, for a higher dimensional system, as a result of a larger number of degrees of freedom in real space, there may be alternatives approaches to reduce the raised potential energy (e.g., deforming the original crystal geometry ) and other consequent structures may exist.

\begin{figure}
  \centering
  \includegraphics[width=3in]{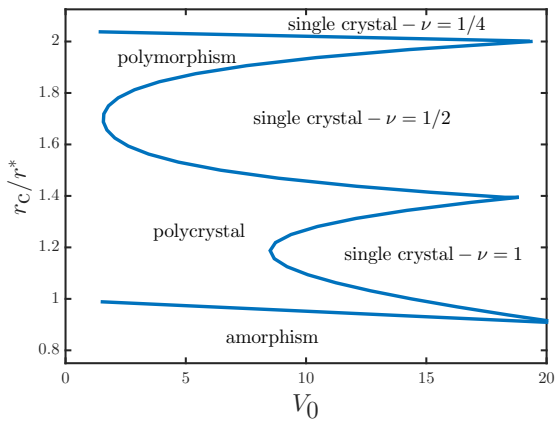}\\
  \caption{Phase diagram of a two-dimensional soft-core ultracold Bose gas on a square lattice $V_{\textrm{latt}}(\mathbf{r})=V_{0}/2 \left[\sin^{2}(\pi x_{1})+\sin^{2}(\pi x_{2})\right]$ versus $r_{\textrm{c}}/r^{\ast}$ and $V_{0}$ demarcated in six distinct crystallographic phases. Following the increase of $r_{\textrm{c}}/r^{\ast}$, the phases are amorphism, polycrystal and polymorphism for a weak confinement, and single crystals with a specific filling factor $\nu$ for a stronger confinement. The interaction strength is fixed at $\alpha=75$.}\label{fig1}
\end{figure}

In this Letter, we use the mean-field Gross--Pitaevskii (GP) equation to demonstrate a variety of crystallographic phases in a Rydberg-dressed Bose gas loaded on a square lattice. By varying the lattice depth and blockade radius, we investigate the crystallographic phase diagram in the absence of any externally imposed frustration, including the lattice geometry and the interaction. In Ref.\cite{Sepúlveda2010}, Sep\'{u}lveda et al. showed that the superfluid fraction depends on the length of the complex network of grain boundaries, and in Ref.\cite{PhysRevA.92.053625}, Lechner et al. proposed a method that allows the tuning of the interaction between vacancies and interstitials by means of external periodic fields. Compared with these two studies, the frustration here is induced only by the constraint of the lattice potential on a quantum elastomer, and the elasticity includes the density modulability and deformable crystal geometry. Compared with the real-space density distribution, the interaction potential can manifest where the lower density is such that we can study the formation of vacancies and interstitials. Most notably, in the present two-dimensional system, we study the anisotropy of these crystallographic phases by calculating the superfluid-fraction tensor\cite{PhysRevB.73.092505,PhysRevB.76.052503}.

For a Rydberg-dressed Bose gas on a triangular lattice\cite{PhysRevLett.116.135303}, the superglass phase is obtained in the absence of externally imposed frustration, e.g., in the lattice geometry or interaction. Here, we study the probable crystallographic structures of a Rydberg-dressed Bose gas on a square lattice by the mean-field method. In the present system, we not only consider the superglass phase, but also obtain other quantum crystallographic phases. Differing from lattice models in which there is a built-in periodic environment, our model is based on the original GP equation with an external periodic potential and an integral kernel that can be viewed as a two-body potential. In the literature, both the simulations of ground states\cite{PhysRevLett.108.265301} and elementary excitations\cite{PhysRevA.87.061602,PhysRevLett.108.175301} by the GP equation and Bogoliubov--de Gennes equations are qualitatively and quantitatively consistent with those that use the path integral quantum Monte Carlo (PIQMC) method, which is a first-principle method. Thus, the superfluid density can be studied in the framework of cold atoms where a mean-field theory can be applied. As a qualitative tool, the mean-field GP method has the advantages of being both a continuous model and easy to calculate.

The two-dimensional GP Hamiltonian for a Rydberg-dressed Bose gas confined into a square lattice is
\begin{equation}\label{hamiltonian}
  \widehat{H}=\frac{-\hbar^{2}\nabla_{\bot}^{2}}{2m}+V_{\textrm{latt}}\left(\mathbf{r}\right)+\Phi\left(\mathbf{r},t\right),
\end{equation}
where $V_{\textrm{latt}}$ is the external square lattice potential with a lattice constant $a$, and $\Phi$ is the interaction potential defined as: $\Phi\left(\mathbf{r},t\right)=\int U\left(\bar{\mathbf{r}}\right)\left|\Psi\left(\mathbf{r'},t\right)\right|^{2}d\mathbf{r'}$
where $U\left(\mathbf{r}\right)$ is the soft-core-interaction kernel and the $\bar{\mathbf{r}}\equiv\mathbf{r}-\mathbf{r'}$ is relative position. Here, the order parameter $\Psi$, which satisfies the normalized condition $\int_{\Omega}\left|\Psi\right|^{2}d\mathbf{r}=1$ ($\Omega$ is a unit cell of the square lattice), is the wavefunction of a Bose--Einstein condensate. In the following, $a$ and $\hbar^{2}/a^{2}m$ are used as units of length and energy, respectively; consequently, the interaction kernel has dimensionless form: $U\left(\mathbf{r}\right)=\alpha/\left(r_{\textrm{c}}^{6}+\mathbf{r}^{6}\right)$, with a tunable strength $\alpha$ and blockade radius $r_{\textrm{c}}$. In general, a contact term should appear in the interaction kernel; however, here, we simply ignore the contact term valid for the case of strong soft-core interaction. This scheme can be performed by using Feshbach resonances, for example. Throughout this paper, we fixed the interaction strength at $\alpha=75$.

Fig.\ref{fig1} shows the phase diagram as a function of the blockade radius $r_{\textrm{c}}/r^{\ast}$, and the lattice depth $V_{\textrm{0}}$. $r^{\ast}$ is defined such that, when $r_{\textrm{c}}=r^{\ast}$, the spontaneous supersolid has a lattice constant equaling $a$. When the external potential is relatively weak, the system undergoes the phases- amorphism (AM), polycrystal (PC), and polymorphism (PM), following the increase of blockade radius. When $V_{\textrm{0}}$ is large enough, the system forms a commensurate structure. To characterize these commensurate structures, (here named single crystals (SCs)) a filling factor $\nu$ is defined as the ratio of the number of occupied and unoccupied sites. As the blockade radius increases, the system undergoes the phases- SC$-\nu=1$, SC$-\nu=1/2$, and SC$-\nu=1/4$. We find that for $r_{\textrm{c}}<r^{\ast}$, an amorphous structure known as superglass (here named amorphism) occurs in an extended region of the phase diagram, and when $r_{\textrm{c}}>r^{\ast}$, stable crystalline structures start to form regionally and compose a PC (with single kind of crystallite) or PM (with more than one kind of crystallite).

\begin{widetext}

\begin{figure}[t]
\begin{center}
\includegraphics[height=3.0in,width=7in]{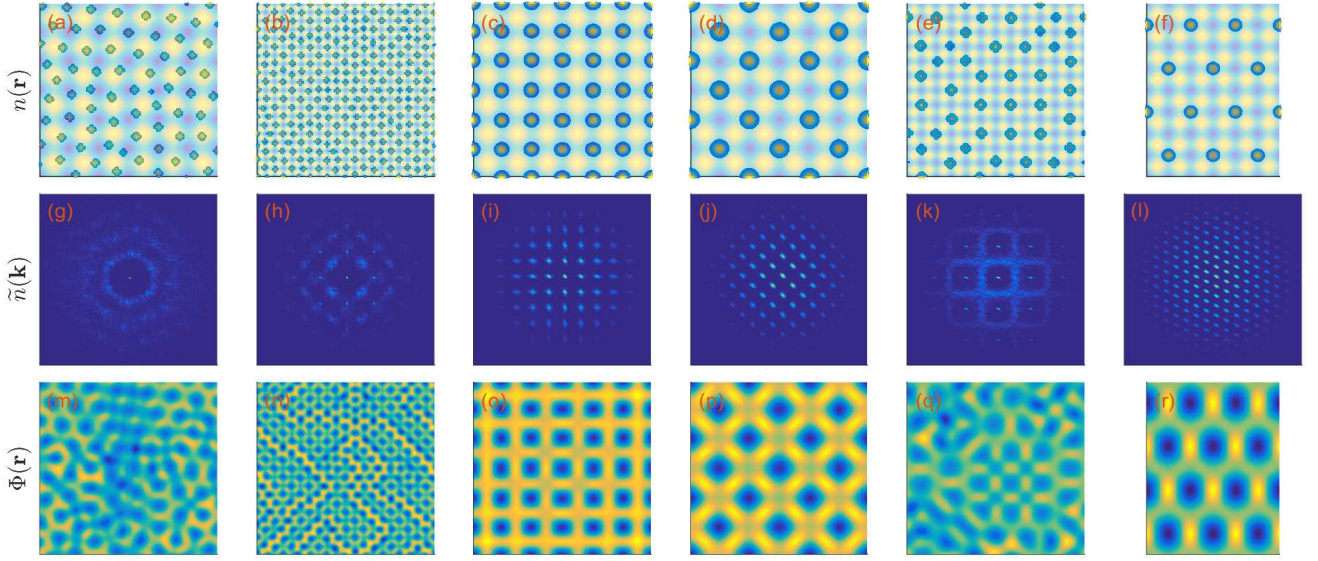}
\caption{Representatives of ground-state density distributions vs. space (a)--(f) and vs. wavevector (g)--(l) for various quantum crystallization states. From left to right, the figures correspond to amorphism ($r_{c}/r^{\ast}=3/4$, $V_{0}=5$), polycrystal ($r_{c}/r^{\ast}=\sqrt{2}$, $V_{0}=5$), single crystal$-\nu=1$ ($r_{c}/r^{\ast}=\sqrt{5}/2$, $V_{0}=10$), single crystal$-\nu=1/2$ ($r_{c}/r^{\ast}=3/2$, $V_{0}=9$), polymorphism ($r_{c}/r^{\ast}=2$, $V_{0}=5$), and single crystal$-\nu=1/4$ ($r_{c}/r^{\ast}=2$, $V_{0}=20$), respectively. (m)--(r) show the interaction potential $\Phi$ for the different quantum crystallization states.}
\label{fig2}%
\end{center}
\end{figure}

\end{widetext}

Fig.\ref{fig2} provides explicit examples of the density $n(\mathbf{r})$ in real space (a)--(f) with the background of lattice potential, and $\widetilde{n}(\mathbf{k})$ in momentum space (g)--(l). $\widetilde{n}(\mathbf{k})$ is the Fourier transform of $n(\mathbf{r})$. Fig.\ref{fig2}(m)--(r) show the interaction potential, $\Phi(\mathbf{r})$, associated with various phases in Fig.\ref{fig1}. In all of the figures, the bright (yellow) color indicates higher values, and the darker (blue) color corresponds to lower values. In Fig.\ref{fig2}(a)--(f), the translucent spots indicate the density droplets.

Fig.\ref{fig2}(a), (g), and (m), Fig.\ref{fig2}(b), (h), and (n), Fig.\ref{fig2}(c), (i), and (o), Fig.\ref{fig2}(d), (j), and (p), Fig.\ref{fig2}(e), (k), and (q), and Fig.\ref{fig2}(f), (l), and (r) correspond to AM, PC, SC$-\nu=1$, SC$-\nu=1/2$, PM, and SC$-\nu=1/4$, respectively. From Fig.\ref{fig2}(a), we find that almost all density droplets avoid the extremes of the potential by distorting its original triangular structure, which results in their random distribution. Fig.\ref{fig2}(g) exhibits the amorphous signature of the density in the momentum space, which distributes in concentric circles. For an AM, Fig.\ref{fig2}(m) shows that the distribution of vacancies and interstitials is also amorphous. Fig.\ref{fig2}(b) shows the ground-state formation composed of many crystallites of varying sizes. The small-dot signals in Fig.\ref{fig2}(h) indicate that there is a single kind of crystallite whose unit cell is square, and the cloudy signals aries from the mismatch between the crystallites. Fig.\ref{fig2}(n) clearly depicts vacancies and interstitials clustering together and forming grain boundaries. Such a phenomenon is similar to the results in Ref.\cite{Sepúlveda2010} and \cite{PhysRevA.92.053625}. For a PM, Fig.\ref{fig2}(e) shows that, in the ground-state formation composed of three kinds of crystallite, one is a square and the other two are quadrature rhombuses. The small-dot signals in Fig.\ref{fig2}(k) indicate the square crystallite, and the cloudy signals are caused by the two quadrature rhombuses. Similar to an AM, Fig.\ref{fig2}(q) shows that the vacancies and interstitials do not cluster together in a PM. It is necessary to classify the quantum crystallographic phase of both the real- and momentum-space distributions. Furthermore, the interaction potential can help us to study the formation of vacancies and interstitials.

The $ij$-th element of the superfluid-fraction tensor $\widehat{f}_{\textrm{s}}\left(\theta\right)$ is defined as
\begin{equation}\label{fs0}
  f_{\textrm{s},ij}\left(\theta\right)=\lim_{q'_{i},q'_{j}\rightarrow0}\frac{m\partial^{2}E^{(1)}\left(\mathbf{q}\right)}{\hbar^{2}\partial q'_{i}\partial q'_{j}},
\end{equation}
where $E^{(1)}\left(\mathbf{q}\right)$ denotes the lowest Bloch band, and $\theta$ is the angle between the quasimomenta $\mathbf{q'}=\left(q'_{1},q'_{2}\right)^{T}$ and $\mathbf{q}=\left(q_{1},q_{2}\right)^{T}$. The Bloch band structures of the system can be obtained by solving the Bloch waves, which are the eigenstates of the nonlinear GP Hamiltonian (\ref{hamiltonian}). The overall
time-dependent wave functions have the following form:
$\Psi\left(\mathbf{r},t\right)=e^{i\mu^{(l)}_{\mathbf{q}}t/\hbar}e^{i\left(\mathbf{q}\cdot\mathbf{r}\right)}\psi^{(l)}_{\mathbf{q}}\left(\mathbf{r}\right)$,
where, for a given wave vector $\mathbf{q}$, $\mu^{(l)}_{\mathbf{q}}$ is the chemical potential.
The corresponding Bloch energy is $E^{(l)}\left(\mathbf{q}\right)=\int\mathcal{E}^{(l)}_{\left(\mathbf{q}\right)}\left(\mathbf{r}\right)d\mathbf{r}$
with the energy density
\begin{multline}\label{Bloch band}
  \mathcal{E}^{(l)}_{\left(\mathbf{q}\right)}\left(\mathbf{r}\right)=\frac{\left|\hbar\left(\nabla_{\bot}+i\mathbf{q}\right)\psi^{(l)}_{\mathbf{q}}\left(\mathbf{r}\right)\right|^{2}}{2m}\\
  +\left[V_{\textrm{latt}}\left(\mathbf{r}\right)+\frac{\Phi^{(l)}_{\mathbf{q}}\left(\mathbf{r}\right)}{2}\right]\left|\psi^{(l)}_{\mathbf{q}}\left(\mathbf{r}\right)\right|^{2},
\end{multline}
and the interaction potential $\Phi^{(l)}_{\mathbf{q}}\left(\mathbf{r}\right)=\int U\left(\bar{\mathbf{r}}\right)\left|\psi^{(l)}_{\mathbf{q}}\left(\mathbf{r'}\right)\right|^{2}d\mathbf{r'}$.
Assuming that the two vectors satisfy the relation $\mathbf{q'}=\widehat{R}\left(\theta\right)\mathbf{q}$ with the two-dimensional rotation matrix $\widehat{R}\left(\theta\right)$, the rotation transformation of the superfluid-fraction tensor can be expressed as $\widehat{f}_{\textrm{s}}\left(\theta\right)=\widehat{R}\left(\theta\right)\widehat{f}_{\textrm{s}}\left(0\right)\widehat{R}\left(\theta\right)^{\dag}$.
The superfluid-fraction tensor $\widehat{f}_{\textrm{s}}\left(\theta\right)$ is diagonalizable, and the eigenvalues of $\widehat{f}_{\textrm{s}}\left(\varphi\right)$ are
\begin{equation}\label{egv}
  \lambda_{\pm}=\frac{\left(f_{\textrm{s},11}+f_{\textrm{s},22}\right)\pm\sqrt{\left(f_{\textrm{s},11}-f_{\textrm{s},22}\right)^{2}+4f_{\textrm{s},12}^{2}}}{2},
\end{equation}
where $\varphi$ is an arbitrary angle.
The first diagonal element of $\widehat{f}_{\textrm{s}}\left(\theta\right)$ is
\begin{multline}\label{fs11theta}
  f_{\textrm{s},11}\left(\theta\right)=f_{\textrm{s},11}\left(0\right)\cos^{2}\left(\theta\right)+f_{\textrm{s},22}\left(0\right)\sin^{2}\left(\theta\right)\\
  +f_{\textrm{s},12}\left(0\right)\sin\left(2\theta\right),
\end{multline}
which may be measurable and can be numerically calculated. When $\mathbf{q}\ll1$, $E^{(1)}\left(\mathbf{q}\right)$ can be
expanded as $E^{(1)}\left(\mathbf{q}\right)-E^{(1)}\left(0\right)\approx\sum_{i,j}f_{\textrm{s},ij}\left(0\right)\left(\hbar^{2}q_{i}q_{j}/2m\right)=f_{\textrm{s},11}\left(\theta\right)(\hbar^{2}{q'}_{1}^{2}/2m)$ by defining $q_{1}=q'_{1}\cos\theta$ and $q_{2}=q'_{1}\sin\theta$, which indicates that the $q'_{1}$ direction is the direction of vector $\mathbf{q}$, and
\begin{equation}\label{fs11}
  f_{\textrm{s},11}\left(\theta\right)\approx\frac{2m\left[E^{(1)}\left(\mathbf{q}\right)-E^{(1)}\left(0\right)\right]}{\hbar^{2}\mathbf{q}^{2}},
\end{equation}
where $\theta=\arctan(q_{2}/q_{1})$.

\begin{figure}
  \centering
  \includegraphics[width=3.4in]{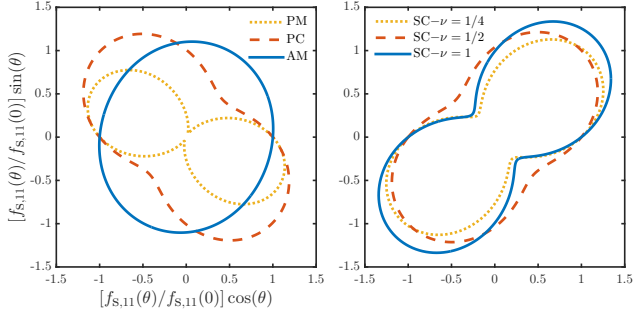}\\
  \caption{The polar plot of $f_{\textrm{s},11}(\theta)/f_{\textrm{s},11}(0)$. Left: PM, PC, and AM phases. Right: SC$-\nu=1$, SC$-\nu=1/2$, and SC$-\nu=1/4$.}\label{fig3}
\end{figure}

Fig.\ref{fig3} shows the polar plots of $f_{\textrm{s},11}$ associated with various phases presented in Fig.\ref{fig2}. A similar technique is presented in Ref.\cite{doi:10.1021/jz500480m}. According to the formulae (\ref{fs11theta}) and (\ref{egv}), the $f_{\textrm{s},11}$ loop should be biconcave unless $f_{\textrm{s},22}\approx f_{\textrm{s},11}$ and $f_{\textrm{s},12}$ is small, i.e., $\lambda_{+}\approx\lambda_{-}$. If $\lambda_{+}=\lambda_{-}$, the superfluid-fraction tensor is reduced to a scalar, and the corresponding system is completely isotropic. For an amorphous structure (e.g., superglass), the $f_{\textrm{s},11}$ loop should be approximately isotropic owing to its randomly distributed density droplets. The concavity--convexity of the $f_{\textrm{s},11}$ loop identifies the isotropy of the system; more precisely, we can define an anisotropy parameter $\eta=\left(\lambda_{+}-\lambda_{-}\right)/\left(\lambda_{+}+\lambda_{-}\right)$. A large $\eta$ indicates large anisotropy. The orientation of the $f_{\textrm{s},11}$ loop indicates the direction of a principal axis, with the other axis being along the vertical direction. For the PC, SC$-\nu=1$, and SC$-\nu=1/2$ phases, the principal axes along $45^{\circ}$ and $135^{\circ}$ are expectable as they are mainly composed of single square cells. In contrast, the orientations of principal axes are not predictable for the AM and PM phases. The orientation of a principal axis is $\left[\arctan(1/2)+\arctan(3/2)\right]/2$ for the SC$-\nu=1/4$ phase.

For the SC$-\nu=1/4$ state, there are two possible configurations: rhombic lattice and square lattice. Which one is the most energetically favorable? To answer this question, we count the bond-leg number of each atom droplet. There are two bonds with length 2, four bonds with length $\sqrt{5}$, and two bonds with length 4 for a droplet of rhombic lattice; in addition, there are four bonds with length 2, and four bonds with length $2\sqrt{2}$ for a droplet of square lattice. As the interaction energy is inversely proportional to the power $\eta$ (here $\eta=6$) of the bond length, we are able to easily estimate the interaction energy. The interaction energy $2/2^{6}+4/\sqrt{5}^{6}+4/\sqrt{13}^{6}=0.065$ of the rhombic lattice is smaller than $4/2^{6}+4/\sqrt{2}^{6}=0.070$ of the square lattice. By this simplified arithmetic, the rhombic lattice is consequently favorable for $\eta=6$. In fact, the rhombic lattice is more energetically favorable only when $\eta>3$, i.e., if the long-rang behavior of interaction is in dipole--dipole form, the square lattice is energetically favorable.
\begin{figure}
  \centering
  \includegraphics[width=3in]{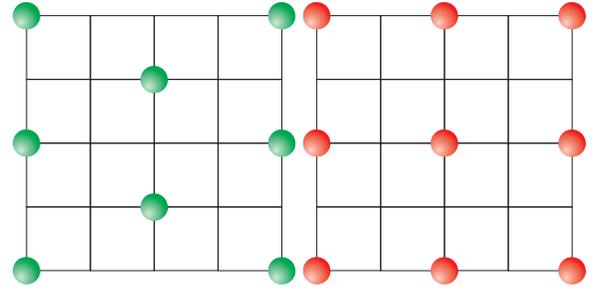}\\
  \caption{Left: rhombic lattice. Right: square lattice.}\label{fig4}
\end{figure}

In this work, numerical simulations using a continuous mean-field model show that quantum crystallographic structures can be investigated in a two-dimensional ultracold atom system loaded on an external periodic potential in the absence of defects. Such a system spontaneously possesses supersolidity originating from a soft-core interaction, and the formation of various structures arises from the mismatch between the supersolid and the external periodic potential. To classify these quantum crystallographic structures, not only the real space density but also the momentum space density and the interaction potential are presented. Most notably, we report a probably measurable quantity on the superfluid characteristic of an anisotropic system. Here, at least qualitatively, we have established a simple but effective model to study quantum crystallography that can be easily generalized to higher dimensional or multicomponent systems, as well as consider additional effects such as synthetic gauge fields or spin-orbit-coupling effects.

This work was supported by JSPS KAKENHI grant numbers JP16H00807 and JP26400366,
and the support from the Ministry of Science and Technology, Taiwan
(under the grant No. MOST 102-2112-M-003-015-MY3)
and the National Center of Theoretical Sciences of Taiwan are acknowledged.

\bibliography{Ref}

\end{document}